\documentclass[namedreferences]{kluwer}
\usepackage{graphicx}

\begin{document}
\begin{article}

\begin{opening}

\title{On Optimality in Auditory Information Processing}

% Author block.
\author{Mattias F.~\surname{Karlsson}\thanks{This work was supported by
        FOA project E6022, Nonlinear Dynamics.}\email{f95-mka@nada.kth.se}}
\author{John~W.C.~\surname{Robinson}$^*$\email{john@sto.foa.se}}
\institute{Defence Research Establishment, SE 172 90 Stockholm, Sweden}

% Abstract and Keywords.
\begin{abstract}
We study limits for the detection and estimation of weak sinusoidal signals
in the primary part of the mammalian auditory system using a stochastic
Fitzhugh-Nagumo (FHN) model and an action-reaction model for synaptic
plasticity.
Our overall model covers the chain from a hair cell to a point just after
the synaptic connection with a cell in the cochlear nucleus.
The information processing performance of the system is evaluated using so
called $\phi$-divergences from statistics which quantify a dissimilarity
between probability measures and are intimately related to a number of
fundamental limits in statistics and information theory (IT).
We show that there exists a set of parameters that can optimize several
important $\phi$-divergences simultaneously and that this set corresponds to
a constant quiescent firing rate (QFR) of the spiral ganglion neuron.
The optimal value of the QFR is frequency dependent but is essentially
independent of the amplitude of the signal (for small amplitudes).
Consequently, optimal processing according to several standard IT criteria
can be accomplished for this model if and only if the parameters are
``tuned'' to values that correspond to one and the same QFR.
This offers a new explanation for the QFR and can provide new insight into the
role played by several other parameters of the peripheral auditory system.
\end{abstract}
\keywords{Auditory system, Information, Detection, Estimation, Divergences}

\end{opening}

\section{Introduction}
When a sensory cell in a mammal is presented with a stimulus, the
information about it must in general be communicated through several layers
of intermediating nerve cells before it reaches the parts of the brain
where the final processing takes place.
A logical question, therefore, is how much of the information is lost in the
first parts of this processing chain and how have these parts of the chain
have (possibly) been optimized by evolution to combat information loss, for
different types of stimuli.
One of the simplest settings of this problem is the auditory system.
The frequency filtering process in the inner ear makes it sufficient in
general, at least for weak signals, to restrict attention to a single type
of stimuli, \textit{a pure tone}, when studying the response of the auditory
nerve cells and their connections in the cochlear nucleus.
From an information-theoretic perspective it is thus of interest to
determine how well the peripheral parts of the auditory processing chain
preserve information about the two parameters, the amplitude and phase, that
characterize a tone at a given frequency.
An even more fundamental question, however, is how well information about the
presence of such a tone is preserved, i.e.\ in what ways this part of the
auditory processing chain imposes limits on achievable detection performance.
Mathematically, these two problems belong to the realm of statistical decision
and information theory (IT); for weak tones the detection problem is,
moreover, intimately connected with the estimation problem of determining the
amplitude.

Despite the extensive literature on information processing in neurons, a
relatively small number of works treat the fundamental statistical limits
for neural detection and estimation that bound the performance of sensory
systems.
One notable exception, however, is Stemmler's work \cite{Stemmler} on the
detection and estimation capabilities of the Hodgkin-Huxley, McCullough-Pitts
and leaky integrate-and-fire model neurons in terms of the Fisher Information.
Stemmler shows that there exists a universal small-signal scaling law which
relates the optimal detection, estimation, and communication performance of
these model neurons, and that this scaling law also applies to the
(narrow-band) signal-to-noise ratio (SNR) on the output of a neuron which is
excited by a sinusoidal signal.
Manwani and Koch \cite{Manwani_Koch} give a detailed analysis of the noise in
dendritic cable structures and its effect on fundamental limits for detection
and estimation.
In particular, they provide relations for minimum mean-square error in linear
estimation and minimum probability of error (the latter under an assumption
of Gaussian noise) based on a stochastic version of the linear one-dimensional
cable equation.
In the majority of other information theoretic analyses of neural information
processing the focus is on the \textit{spike train} on the output of a neuron
though, and a long-standing objective has been to try to break the ``neural
code'' of the spike train.
However, there is a fundamental component missing in modeling that rests
solely on considering information in the spike train and it is the
influence of the \textit{synaptic connections}.
The importance of this aspect of neural computation has recently been
recognized and it has even been suggested that the synaptic connections in
fact represent the primary bottleneck that limits information transmission
in neural circuitry \cite{Zador}.
Consequently, when studying information processing in neurons, in particular
detection and estimation capabilities of the auditory system, it seems
imperative to consider models and methods that describe not only the
individual neurons and their spike trains but also the synaptic connections
between the neurons.

In the present study we investigate, theoretically, the fundamental limits
for detection and estimation of weak signals in the mammalian auditory system.
We model the neurons in the auditory nerve and their synaptic connections 
using ideas from Tuckwell \cite{Tuckwell} and Kistler-Van~Hemmen
\cite{Kistler_VanHemmen} that take into account the notion of
\textit{synaptic plasticity}.
Incorporation of the synaptic efficacy's dependence on the prehistory of
action potentials arriving to the synapse in the model makes it possible to
obtain a more realistic assessment of the information available to the next
step in the auditory processing chain, the processing in the cochlear nucleus.
Another feature of our study is the use of more general measures of
signal-noise separation.
To quantify signal-noise separation we use the so-called $\phi$-divergences
from statistics and IT \cite{Liese_Vajda}.
The $\phi$-divergences are applicable to virtually any kind of signal and
system (in a stochastic setting), in particular the highly nonlinear dynamic
systems represented by neurons, and are intimately related to a number of
fundamental limits in statistics/IT.
Our main objective is to determine whether the primary auditory system has a
structure whereby (nontrivial) optimizations of $\phi$-divergences with
respect to parameters can occur.
Given the significance of the $\phi$-divergences as performance measures, an
affirmative answer to this question would yield a new view on the role
played by various parameters in the neurons of the auditory system, such as
the quiescent firing rate (QFR), and would inspire new experiments relating to
the function of the auditory processing chain.
We show that such optimizations indeed are possible, where some of the
underlying mechanisms are explained in terms of the model structure, and we
numerically determine the optimal values.

The paper is organized as follows.
In Section \ref{Methods} we describe our model of the auditory system,
in which the central component are the Fitzhugh-Nagumo equations.
This section also includes an introduction to $\phi$-divergences and a review
of their properties.
The divergences are computed in Section \ref{Results}, and discussed in
Section \ref{Discussion}.
\section{Methods}
\label{Methods}
\subsection{Physiological modeling}
We consider the peripheral part of the mammalian auditory nervous system
\cite{Geisler}, beginning with the acoustic (fluid) pressure at a point in
the inner ear and ending at the soma of a cell in the cochlear nucleus.
As a model of the chain from the inner ear, via an inner hair cell and a 
spiral ganglion cell, to a point a small distance down the ganglion axon we
employ a stochastic FitzHugh-Nagumo (FHN) model \cite{FitzHugh,Scott}.
This model, which we henceforth (with a slight abuse of language) will call
the FHN neuron, represents an attractive choice in our study for two reasons:
It is analytically/numerically tractable and has the ability to produce a
response that is both visually and statistically similar to that observed in
real neurons.
In particular, it is well-known that even simple (white-noise driven)
stochastic FHN models are able to accurately reproduce the interspike
interval histograms (ISIH) in various forms of nerve fibers, such as the
auditory nerve fibers of squirrel monkeys \cite{Massanes_Vicente99}.
For the terminal boutonic connections of the auditory nerve with the dendrites
(or soma) of the cells in the cochlear nucleus, together with the parts of 
the dendrites from the boutonic connections to the somas, we employ an
action-reaction model combined with a time-varying $\alpha$-function like
transformation with additive noise \cite{Tuckwell,Kistler_VanHemmen}.
The conjunction of these two model features makes it possible to capture
both the synaptic plasticity and variability observed in real neurons.
Furthermore, incorporation of plasticity in the model turns out to be of
crucial importance for our results since it removes ``false optima'' that
would otherwise be present.
\subsubsection{Stochastic FitzHugh-Nagumo Model.}
\begin{figure}[ht]
\begin{center}
\includegraphics[width=8cm, height=7cm]{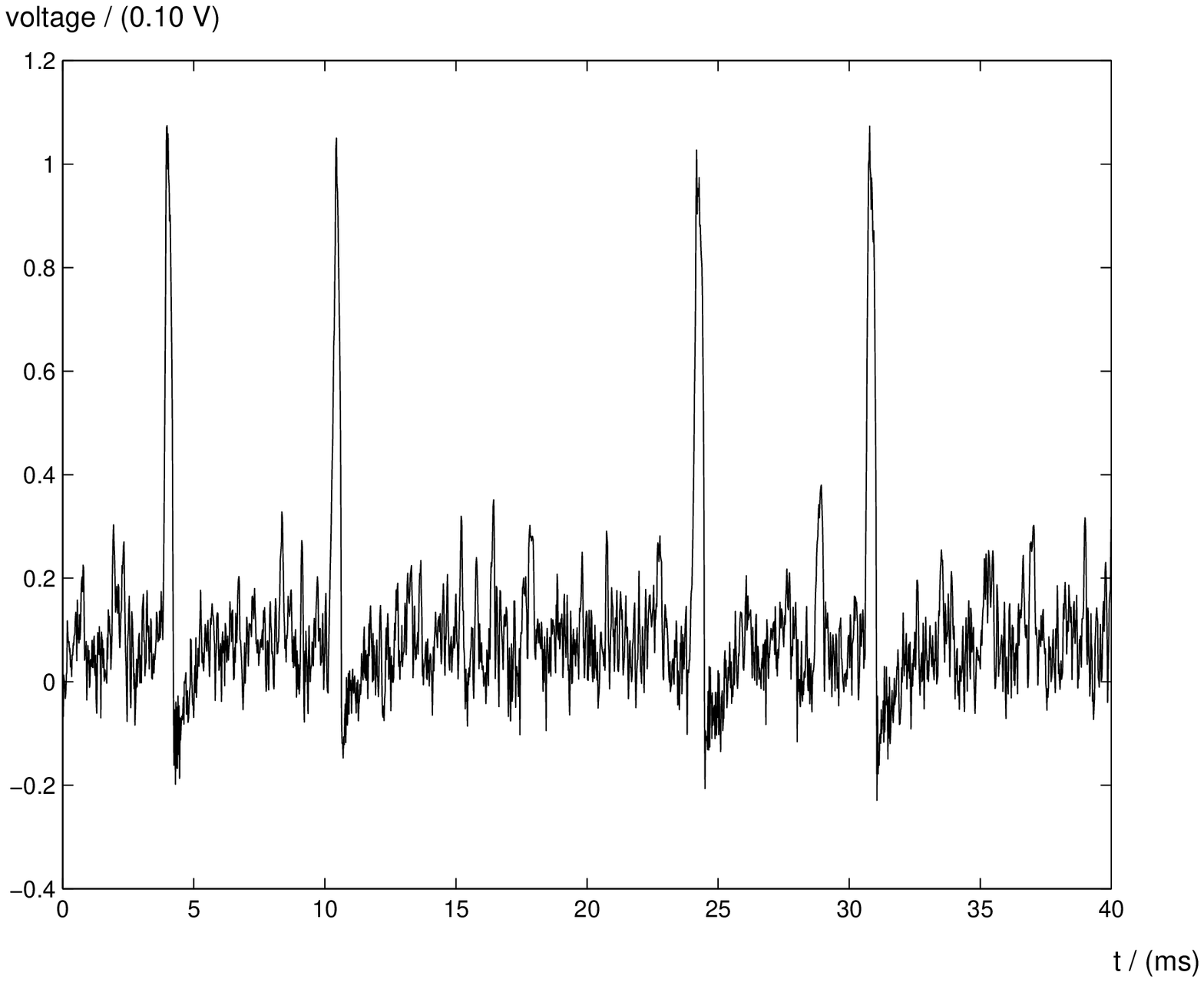}
\caption{A typical example of the output from the model
         (\ref{FHN}),(\ref{OU_process}) with parameters $a=0.55$, $b=0.12$, 
         $\delta=1$, $\varepsilon = 0.005$, 
         $\sigma=100\sqrt{2\cdot 10^{-5}}$ and $\lambda=100$, when an input
         signal $s_t$ as in (\ref{sinusoid}) with parameters $A=0.1$ and
         $\omega_0=8$ is applied.}
\label{FHN_output}
\end{center}
\end{figure}

The \textit{stochastic FHN model} is given by the following system of
stochastic differential equations \cite{Longtin}
\footnote{To guarantee global solutions to (\ref{FHN}) we must assume that
          the model for very large $|V_t|$ is modified so that the potential
          in $V_t$ grows at most linearly.} 
\begin{equation}
 \begin{array}{rcl}
  \varepsilon dV_t & = & V_t(V_t -a)(1-V_t) \, dt - W_t \, dt + dv_t, \\
              dW_t & = & (V_t - \delta W_t - (b + s_t)) \, dt, \\
 \end{array}, \quad t \in [0,T],
\label{FHN}
\end{equation}
where $\varepsilon,a,b,\delta > 0$ are (nonrandom) parameters, $V$ is the
fast (``voltage like'') variable, $W$ is the slow (``recovery like'')
variable, and $s_t$ is the signal process representing the \textit{stimuli},
here the acoustic pressure in the inner ear.
The parameter $a$ effectively controls the barrier height between the two
potential wells in the potential term (i.e., the first term on the RHS of the
first equation) and the variable $b$ is a bias parameter moderating the effect
of the signal input.
These two parameters affect the \textit{stability} properties of the FHN
neuron, and so does the relaxation parameter $\delta$ multiplying the slow
variable.
The parameter $\varepsilon$ sets the \textit{time scale} for the motion in
the potential described by the first equation.
Normally, the variable $V$ is thought to represent \textit{membrane voltage}
in the neuron but since the FHN model can be viewed as obtained by ``descent''
from the higher dimensional Hodgkin-Huxley model (or other likewise more
elaborated models) it is not reasonable to attach a too strict physical
meaning to it.
To us it will merely act as a convenient way of modeling the timing
information in the action potentials generated by the neuron when the latter
are defined via a simple threshold operation on the fast variable $V$.
The signal $s_t$ is here chosen to enter on the slow variable $W$, which
controls the \textit{refractory periods} of $V$, in order to facilitate a
comparison with the qualitative results for the corresponding deterministic
dynamics in \cite{Alexander_et_al}.
However, it is easy to transform the system into an equivalent one (of the
same form) where the signal enters on the fast variable
\cite{Alexander_et_al}.
The stochastic process $v_t$ is a noise process accounting for the
\textit{variability in firing pattern} observed in real neurons, which we
in order to have control over the correlation time \cite{Longtin} take to 
be an Ornstein-Uhlenbeck (OU) process
\begin{equation}
 dv_t = -\lambda v_t \, dt + \sigma d\xi_t, \quad t \in [0,T],
\label{OU_process}
\end{equation}
where $\lambda > 0$ determines the effective correlation time and $\sigma > 0$
is the intensity of a standard Wiener process (integrated Gaussian white noise)
$\xi$.
We assume that all the input and intrinsic noise sources can be collectively
described by this process.
This noise model is also often used with $\lambda=0$, so that $v_t$ becomes
a Wiener process, which has proved sufficient to reproduce real data, see
e.g.~\cite{Massanes_Vicente99}.
An example of an output to the FHN neuron (\ref{FHN}),(\ref{OU_process}) with
sinusoidal signal and parameter values typical for the simulations is shown
in Fig.~\ref{FHN_output}.
\subsubsection{Spike train.}
An important underlying assumption in our model and, indeed, in most
rate-based treatments of neural dynamics, is that the \textit{intervals}
between action potentials, not their particular form, in a given neuron carry
all the information relevant to the subsequent neural processing by other
connected neurons.
Accordingly, in the remaining parts of the model that describe how the output
of the FHN neuron is processed we replace the output of the FHN neuron by an
equivalent random point process
$$
 \mathcal{T} =
 \{ 0 < \tau_0 < \dots < \tau_{k-1} < \tau_k < \tau_{k+1} \dots \leq T\},
$$
(the number of points in $\mathcal{T}$ may be finite or infinite) where
$\tau_k$ is defined by level crossings of the fast variable $V$ in the FHN
model as
$$
 \tau_{k+1} = \inf \{t > \tau_k: V_t > \gamma \mbox{ and } V_s < \gamma
              \mbox{ for some } \tau_k \leq s \leq t\}.
$$
In other words, $\tau_{k+1}$ is the first time after $\tau_k$ for an
upcrossing over the level $\gamma$ ($\tau_{0}$ is the first time
for an upcrossing after $t=0$), where $\gamma$ is suitably chosen to
represent an \textit{action potential level}.
The point process $\mathcal{T}$ thus contains the timing information in the
nerve signals at a point in the auditory nerve immediately after the ganglion
cell and will therefore be referred to as the \textit{spike train}.
Since the shapes and relative positions of the action potentials are not
appreciably changed as they propagate through the (myelinated) axons of the
auditory nerve we assume that the process $\mathcal{T}$ also represents the
timing information in the action potentials as they reach a terminal
connection in the telodendria of the ganglion cell.
\footnote{The time delay incurred by the propagation down along the auditory
          nerve will be neglected since it will be approximately 3-4\% of
          the length $T$ of the observation time interval in our examples.}
\subsubsection{Synaptic connections.}
The model for synaptic response is made up of two parts; a nominal (or
average) response and a variability from the nominal due to synaptic
plasticity \cite[ch.\ 13]{Koch}.

For a synapse in a nominal state at an electrotonic distance $x_0$ from the
soma on a dendrite of some length $L \geq x_0$, the impulse response $r$
(``Green's function'') for the transformation from action potential applied
on the presynaptic side of the synapse to the voltage at the soma can be
modeled by an expansion of the form \cite[sec.\ 6.5]{Tuckwell}
\begin{equation}
 r(t) =
 \beta \sum_{n=0}^\infty \frac{A_n(x_0)}{1 + \lambda_n^2 - \alpha}
      \Big( t e^{-\alpha t} - \frac{ e^{-\alpha t} + e^{-t(1 + \lambda_n^2)}
            }{ 1 + \lambda_n^2 - \alpha} \Big),
 \quad t \geq 0,
\label{super_alpha}
\end{equation}
(with uniform convergence) where $r(t) = 0$ for $t < 0$.
Expressions for the constants $A_n,\lambda_n$ in terms of $L$, and graphs
showing the appearance of (\ref{super_alpha}) for typical values of these
constants and $\alpha,\beta$, are given in \cite{Tuckwell}.
In (\ref{super_alpha}) it is assumed that the impulse response from action
potential to post-synaptic current at the soma is given by a so-called
$\alpha$-function of the form $h(t) = \beta t e^{-t \alpha}$ for $t  \geq 0$
and $h(t) = 0$ for $t < 0$ \cite{Jack_Redman}.
From the definition of $r$ it is clear that the expression (\ref{super_alpha})
actually describes both the synapse and a part of the dendrite (the part
between the synapse and the soma), but since the response at a point down the
dendrite is mainly determined by the response of the synapse we shall, for
simplicity, refer to $r$ in (\ref{super_alpha}) as the \textit{nominal
synaptic response}.

The synaptic connections in the cochlear nucleus are often made by synapses
having a fair, or even a large amount, of release sites, such as the
\textit{endbulb of Held}, which is connected to spherical bushy cells in the
anteroventral cochlear nucleus \cite{Webster_et_al}.
As a consequence, the synaptic transmission will be reliable in the sense
that an incoming action potential will almost always yield an
\textit{excitatory postsynaptic potential} (EPSP).
However, the EPSPs will vary in strength depending (primarily) on the
prehistory of the action potentials that have arrived at the synapse.
This phenomenon, the \textit{synaptic plasticity}, has a crucial effect on
the overall dynamical behavior of the nerve and needs to be taken into
account in conjunction with the nominal response in (\ref{super_alpha}).
We model the plasticity using a simple action-recovery scheme developed by
Kistler and Van Hemmen \cite{Kistler_VanHemmen} which combines the three
state plasticity model of Tsodyks and Markham \cite{Tsodyks_Markham} and the
spike response model of Gerstner and van Hemmen \cite{Gerstner_VanHemmen}.
The action-recovery scheme employs a variable $Z$ and its complement $1-Z$
that correspond to ``active'' and ``inactive'' \textit{resources},
respectively, where the term ``resources'' can be interpreted as resources on
both the pre- and the postsynaptic side, such as the availability of
neurotransmitter substance or postsynaptic receptors. 
In addition, resources can also be interpreted as some ionic concentration 
gradient, e.g. the membrane potential on the postsynaptic side.
This approach, therefore, also compensates for the EPSPs' dependence on the
voltage of the following neuron's soma.
Quantitatively, the amount of available resources are determined by the
recursion \cite{Kistler_VanHemmen}
\begin{equation}
 Z_{\tau_k+1} =
 1 - [1 - (1-R)Z_{\tau_k}] \exp[-(\tau_{k+1} - \tau_k) / \tau],
\label{R_and_Z}
\end{equation}
where $0 < R \leq 1$ is a constant corresponding to the fraction of resources 
that gets inactive due to a spike and $\tau > 0$ is a decay time parameter. 
The variable $Z_{\tau_k}$ should be interpreted as the amount of resources
available just before time $\tau_k$ and it is therefore proportional to the
strength in an eventual EPSP caused by an action potential arriving to
the synapse at time $\tau_k$.
An approximation to the initial condition $Z_{\tau_0}$ can be obtained by
forming an average of the available resources for a number of spike trains,
generated  by the unforced FHN model for the studied system, for a large $T$.
Thus, by using the plasticity model above we can calculate the
\textit{pristine} (or noise-free) \textit{postsynaptic response} $R$ as
$$
 R(t,x) = \sum_{k=0}^\infty Z_{\tau_k} r(t-\tau_k), \quad t > 0,
$$
where $r$ is the nominal response given in (\ref{super_alpha}).
This model is capable of producing results in close agreement with real data
(cf.\ \cite{Tsodyks_Markham}), provided the appropriate choices of constants
are made.

In reality there is always also a certain noise present due to
e.g.\ the inherent unreliability of the ionic channels involved in the
transmission of signals in and between the neurons \cite{Koch}.
To take this effect into account we have added zero mean white Gaussian noise
with intensity $\sigma_2$ to the EPSPs given by our model, which thus
represents our \textit{total synaptic response}.
\subsection{Information processing}
\label{Inf_proc}
We study information processing performance in terms of general statistical
signal-noise separation measures applied to the output of our model, the
soma of a cell in the cochlear nucleus.
The output signal-noise separation setting was chosen since it can be applied
with only minimal assumptions about the input signal.
Due to the frequency selectivity of the primary parts of the auditory system
it is sufficient, at least as a good first approximation for weak signals
\cite{Eguiluz_et_al,Camalet_et_al}, to restrict attention to sinusoidal
signals (possibly with slowly varying amplitude and phase).
The simplicity that the output-separation setting offers can be contrasted
with that of a communications setting which in general would require
considerably more assumptions in order to define quantities like alphabet,
message,
\footnote{The message set involved (at each given frequency), if one can be
          defined, would depend entirely on the situation; it would be
          different for various phrases in human languages and would
          be different for natural sounds in different environments.
          This makes it reasonable to assume that the primary parts of the
          auditory system have been optimized by evolution with respect to
          criteria that are largely invariant, such as the ability to
          detect and possibly determine the amplitude of a weak tone.}
coding and channel capacity \cite{Cover_Thomas}.
Of course, one could also select some stochastic signal and consider only
mutual information between input and output but this too would require some
further statistical assumptions.
For our study however, it is sufficient to restrict attention to the very
simple class of signals $s_t$ in the FHN model of the form
\begin{equation}
 s_t = A \sin(\omega_0 t + \varphi),
\label{sinusoid}
\end{equation}
where $A, \omega \geq 0$ are constant in time and $\varphi \in [-\pi,\pi)$
is a phase which is also constant in time.
\subsubsection{$\phi$-divergences and generalized SNR.}
\label{diverg_sect}
A number of fundamental limits in statistical inference and IT can be
expressed as monotonic functions of so-called $\phi$-divergences, which can
be though of as ``directed distances'' between probability measures.
For example, the minimal probability of error in (Bayesian) detection,
Wald's inequalities (sequential detection), the bound in Stein's lemma
(cutoff rates in Neyman-Pearson detection) and the Fisher information for
small parameter deviations (the Cram\'er-Rao bound) can all be written as
simple functions
of a $\phi$-divergence.
In the simplest setting, where $p_0,p_1$ are two probability density
functions (PDFs) on the real line $\mathbb{R}$, the
$\phi$-\textit{divergence} $d_\phi(p_0,p_1)$ between $p_0,p_1$ is defined as
\cite{Liese_Vajda}
\begin{equation}
 d_\phi(p_0,p_1) =
 \int_\mathbb{R} \phi \Big ( \frac{p_1(x)}{p_0(x)} \Big) p_0(x) \, dx,
\label{divergence definition}
\end{equation}
where $\phi$ is any continuous convex function $\phi$ on $[0,\infty)$ (we
assume $p_0(x)=0$ if $p_1(x)=0$).
A $\phi$-divergence satisfies $d_\phi(p_0,p_1) \geq 0$, with equality if and
only if $p_0 = p_1$ almost everywhere, and thus expresses the ``separation''
between $p_0,p_1$ in a relative-entropy like way.
Indeed, one prominent member of the family of $\phi$-divergences is the
Kullback-Liebler divergence or relative entropy, also known as
\textit{information divergence} $d_I$ \cite{Cover_Thomas}, obtained for
$\phi(x) = - \ln(x)$.
Other important members of this family are the \textit{Kolmogorov} or error
divergence $d_\mathcal{E}^{(q)}$, obtained for $\phi(x) = |(1-q)x - q|$ where
$q\in [0,1]$ is a parameter, and the $\chi^2$-\textit{divergence}
$d_{\chi^2}$, obtained for $\phi(x) = (1-x)^2$.

The $\chi^2$-divergence is twice the first term in a formal expansion of the
information divergence around 0 (i.e.\ for $p_0 = p_1$) and is a (tight)
upper bound for a family of generalized SNR measures known as deflection
ratios
\footnote{Indeed, it can be shown that the (narrow-band) SNR measures used
          in stochastic resonance can be expressed as limits of deflection
          ratios \cite{Rung_Robinson,we_and_us}.}
that depend only on the means and variances of the observables.
If $h$ is some function of data, the \textit{deflection ratio} (DR) $D(h)$ is
defined as
\cite{Basseville}
$$
 D(h) = \frac{ |E_1(h) - E_0(h)|^2 }{ \mbox{Var}_0(h) },
$$
where $E_1(h),E_0(h)$ is the expectation of $h$ computed using $p_0$ and
$p_1$, respectively, and $\mbox{Var}_0(h)$ is the variance of $h$ computed
using $p_0$.
The DR is upper-bounded as
\begin{equation}
 D(h) \leq d_{\chi^2}(p_0,p_1),
\label{defl_bound}
\end{equation}
with equality if and only if $C_1(h - E_0(h)) = C_2 (p_1/p_0 -1)$ with
$p_0$-probability one, for two constants $C_1,C_2$ not both zero.
In particular, we have equality in (\ref{defl_bound}) if $h$ equals
$p_1/p_0$, the \textit{likelihood ratio}.
It follows that a larger $\chi^2$-divergence allows for larger SNR, when
expressed in terms of DRs.

The $\chi^2$ and information divergences determine locally the Cram\'er-Rao
bound (CRB) for parameter estimation (\cite{Salicru,Cover_Thomas}).
For example, if $\theta$ is a parameter with values in some open interval
$\mathcal{I}$ and $p_\theta, \theta \in \mathcal{I},$ is a family of PDFs on
$\mathbb{R}$ indexed by $\theta$ then, under some regularity conditions,
$$
 \lim_{\theta \rightarrow \theta_0}
 \frac{d_I(p_{\theta_0},p_\theta)}{(\theta - \theta_0)^2}=
 \lim_{\theta \rightarrow \theta_0}
 \frac{d_{\chi^2}(p_{\theta_0},p_\theta)}{2(\theta - \theta_0)^2} =
 \frac{1}{2}I(\theta_0),
$$
for $\theta_0 \in \mathcal{I}$, where $I(\theta_0)$ is the Fisher Information
at $\theta_0$.
Thus, for estimation of $\theta$ when $\theta$ is near $\theta_0$ the CRB
(which is the inverse of the Fisher Information), and thereby the achievable
accuracy for unbiased estimation of $\theta$, is locally determined by the
growth of the $\chi^2$ and information divergences as a functions of
$\theta$, near $\theta_0$.

The Kolmogorov divergence is directly related to the minimal achievable
probability of error in Bayesian hypothesis testing.
If $p_0$ and $p_1$ are two possible PDFs for the data observed and $q$ is
taken as the \textit{a priori} probability of $p_0$ to be correct, so that
$p_1$ has probability $1-q$, then the minimal achievable probability of
error
\footnote{As is well-known, $\tilde{P}_e^{(q)}(p_0,p_1)$ is achieved
          with a simple likelihood ratio test.}
$\tilde{P}_e^{(q)}(p_0,p_1)$ for decision between $p_0,p_1$ (i.e.\ which is
the correct density) based on a single sample $x$ is given by
(cf.\ e.g.~\cite{Ali_Silvey})
$$
 \tilde{P}_e^{(q)}(p_0,p_1) =
 \frac{1}{2} \big( 1 - d_\mathcal{E}^{(q)}(p_0,p_1) \big).
$$
A larger Kolmogorov divergence thus gives a smaller minimal probability of
error.

For later reference we point out that all the definitions and properties
above have counterparts on much more general probability spaces
\cite{Liese_Vajda,we_and_us,Rung_Robinson}, for instance in the
infinite dimensional context of probability measures on the space of
continuous functions on $[0,T]$.
\subsubsection{Auditory processing performance.}
\label{aud_proc_perf}
In order to apply $\phi$-divergences to assess performance in our model of
the auditory processing chain, we need to specify the setting in somewhat 
greater detail, as well as elaborate some of the features of the model.

We have chosen to make the parameters $A,\omega$ and $\varphi$ constant,
which in a detection scenario means that we are considering so-called
coherent detection (detection of a completely deterministic signal).
At first this might seem as an oversimplification but we argue that it is
not, for the following reason.
There are a number of nerve cells in the auditory nerve ``tuned'' to any
given frequency and each corresponding axon, moreover, exhibits spatial
divergence near the end where it splits up into different branches.
Connections are then made between these branches and the dendritic tree 
or soma of the following neurons.
Since the dendrites (from the connective synapse to the soma) have different
lengths, the time delays in them will be different.
For sinusoidal input signals this can be exchanged for a phase shift of the
signal, at least as a good first approximation.
Thus, for a given frequency, the primary auditory processing can be viewed as
taking place over a ``bank'' of parallel channels, similar in
characteristics but corresponding to different phase shifts.
In a detection setting this corresponds to a bank of coherent detectors
operating on the outputs of these channels.
It is conceivable that the subsequent processing can take advantage of this
low-level parallelism and that detection is possible based on a logical
``or'' operation where one detector indicating presence of the signal is
sufficient. 
Therefore, we use a fixed phase $\varphi$ in the signal $s_t$ in
(\ref{FHN}),(\ref{sinusoid}) and treat the phase as a (variable) parameter.

\begin{figure}
\center
  \begin{tabular}{c}
  \includegraphics[width=7cm,height=7cm]{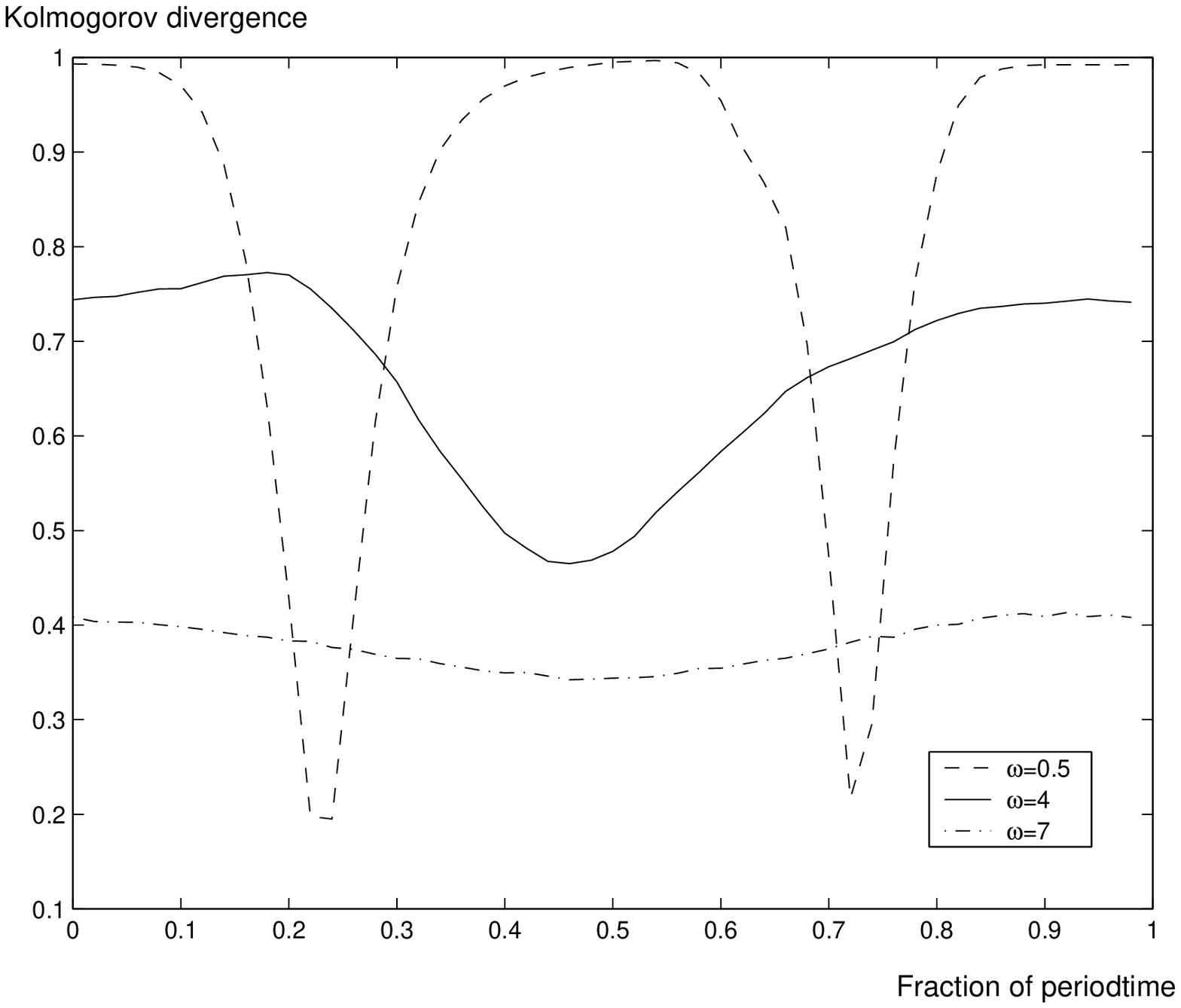} \\
  \includegraphics[width=7cm,height=7cm]{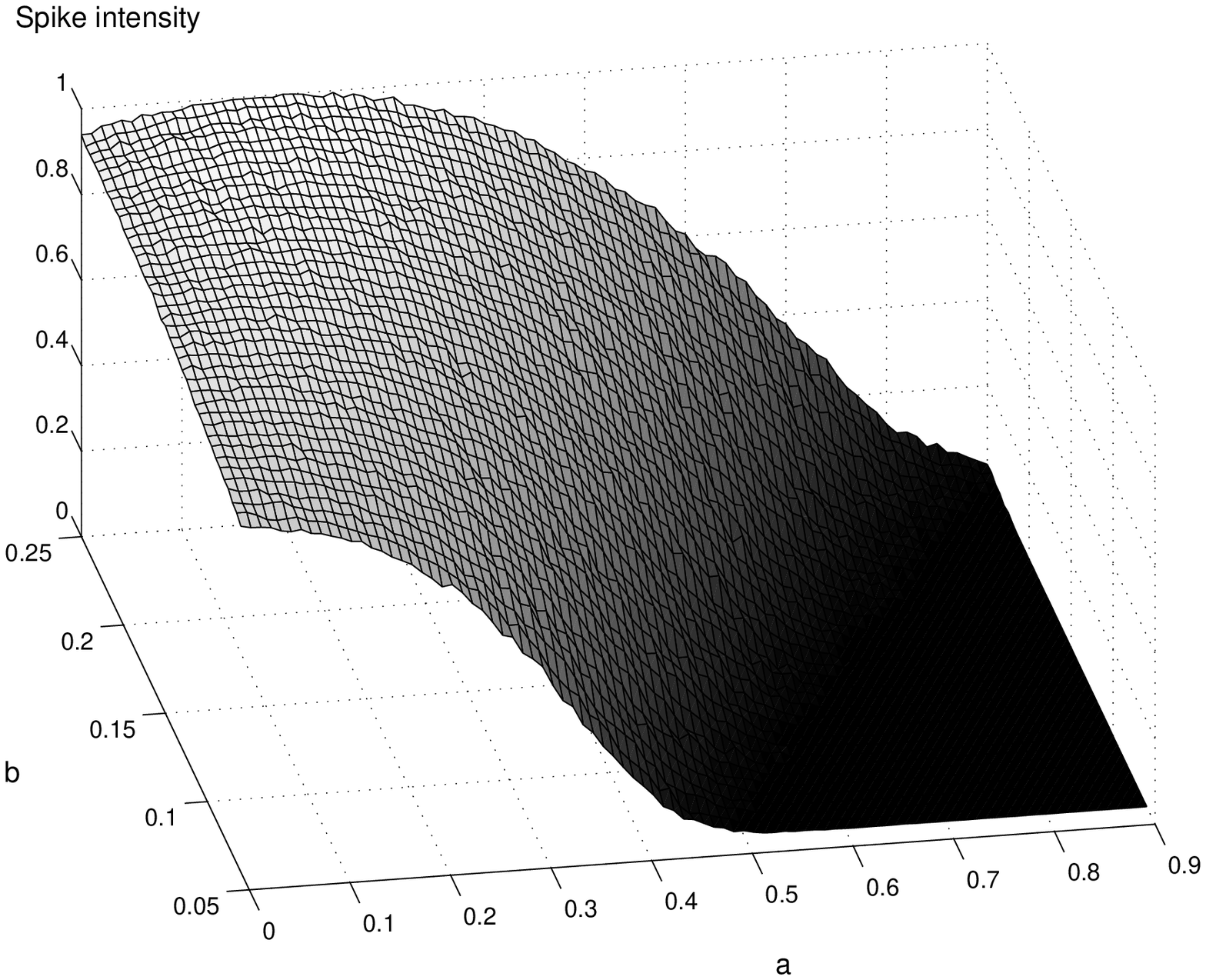}
  \end{tabular}
\caption{(a) Top: Kolmogorov divergence variation for $q=0.5$ on the output
             of the model over one period of the signal $s_t$ in
             (\ref{sinusoid}) for three different frequencies and amplitude
             $A=0.2$.
             The parameters in the FHN model (\ref{FHN}) are $a=0.3$,
             $b=0.12$, $\delta=1$ and $\varepsilon=0.005$ and the other
             parameters are $\sigma=100\sqrt{2\cdot 10^{-5}}$, $\lambda=100$,
             $\alpha=10$, $\beta=100$, $x_0=0.25$, $L=1.5$, $R=0.2$,
             $\tau=50$, $\gamma=0.5$.
         (b) Bottom: Spontaneous activity (no input signal applied) for the
             FHN-model, for different values of the parameters $a$ and $b$
             and with the other  parameters set to $\delta=1$,
             $\varepsilon = 0.005$, $\sigma=100\sqrt{2\cdot 10^{-5}}$ and
             $\lambda=100$.}
\label{two_fig}
\end{figure}

We assess the auditory processing performance by computing the
$\phi$-divergences
of the output of our model (the voltage to the soma of a cell in the cochlear
nucleus) at a time point $T$, where $T$ is the end point of a long time
interval $[0,T]$.
The two PDFs $p_0,p_1$ in the definition (\ref{divergence definition}) are in
the present setting given by the PDF for the output when no signal is present
in the FHN model (\ref{FHN}),(\ref{OU_process}) (i.e.\ $s_t \equiv 0$) and
when a signal $s_t$ as in (\ref{sinusoid}) is present, respectively.
Since the PDFs in this case are densities on the real line they are easy to
compute, using numerical simulation, but they are dependent on the phase
$\varphi$, and so are the resulting $\phi$-divergences.
In order to overcome this, and obtain overall performance measures of the
processing over all the parallell channels described above, we have weighted
together the $\phi$-divergences as
\begin{equation}
 d_\phi = \int_0^{2 \pi} d_\phi(\varphi) \, d\varphi,
\label{av_div}
\end{equation}
based on the assumption that there are enough channels to cover a
sufficiently dense set of the phase interval $[0,2 \pi)$.
It turns out that for high frequencies the $\phi$-divergences do not vary
appreciably over a period but in the medium and low frequency cases there
will typically be one or two regions of phase values where the
divergences are significantly lower, as illustrated in Fig.~\ref{two_fig}a.
However, since the regions where the $\phi$-divergences deviate significantly
from their average values generally are relatively small we argue that average
divergences as in (\ref{av_div}) are relevant as measures of system
performance.

Finally we remark that even though the more general ``infinite dimensional''
formulas for divergences mentioned above in principle could be applied if
we generalized the problem to the case where output over a whole time
interval $[0,T]$ was observed (instead of only its end point $T$), these
formulas are considerably more difficult to handle numerically and involve
solving a general nonlinear filtering problem \cite{Liptser_Shiryayev}.
Since the synaptic connection itself represents an averaging over time (and
thus ``dimensionality reduction'' in the problem) we have chosen the
approach above as a reasonable compromise to reduce computational complexity
while retaining relevance of the model.
\subsection{Simulations}
The stochastic differential equations were solved using the Euler-Maru\-yama
scheme \cite{Kloeden_Platen} and the PDFs of the output to the model were
estimated using a histogram approach based on counting the number of samples
falling in a grid of intervals on the real line.
For calculation of the Kolmogorov divergence the so obtained ``raw''
histograms were sufficient but they proved insufficient for the $\chi^2$ and
information divergences (which are sensitive to inaccuracies in the
representation of the PDFs).
Therefore, smoothing with a kernel of the type $e^{-c|x|}$ was applied to the
estimated PDFs before the latter two divergences were calculated.
In order to reduce the dependence on the smoothing parameter $c$, its values
were kept in a region where the results for the Kolmogorov divergence did not
vary appreciably depending on weather smoothing was applied or not.
Moreover, in this region, the values of the so computed $\chi^2$ and
information divergences were qualitatively independent of the value of $c$.
All our simulations were done using Matlab on UNIX(Digital)/Linux(i386).
\section{Results}
\label{Results}
Our main object of study is the variability of performance, quantified via
$\phi$-divergences (cf.\ Sect.~\ref{aud_proc_perf}), as a function of
parameters.
We shall primarily focus on the Kolmogorov divergence, since this is easiest
to compute numerically, but we shall also consider performance in terms of
the information and $\chi^2$ divergences, and deflection ratios.

The regimes of values used for the parameters in the FHN part of the model 
are chosen on the basis of previous studies 
\cite{Massanes_Vicente99,Alexander_et_al}.
First, a nominal set of parameters is chosen for which the FHN output 
resembles real neuron data and then the parameters are varied around 
this point.
At all times, however, the parameters are kept inside the region where the
output is spike-train like i.e., all the resulting FHN outputs are visually
similar to the one shown in Fig.~\ref{FHN_output}.
The synaptic constants used for the simulation are chosen in order to 
give realistic EPSP:s for the studied systems and the distance $x_0$ is set 
rather small ($x_0=0.25$ on a dendrite of length $L=1.5$) since many 
synapses in the auditory system, e.g.\ the endbulb of Held, form 
connections close to the soma.
\subsection{Performance with respect to variation of $a$ and $b$.}
\label{Perf_a_b}
A basic example of performance expressed as a function of parameters is
shown in Fig.~\ref{a-b_variations}a where the Kolmogorov divergence for 
$A=0.2$, $\omega_0=8$, $\delta=1$ and $\sigma=100\sqrt{2\cdot 10^{-5}}$ 
is displayed as a function of $a$ and $b$.
Both $a$ and $b$  have effect on how much excitation that is needed to
produce spikes in the FHN output.
If $a$ is made smaller the potential barrier height decreases, which gives
a larger spike rate.
Increasing the value of $b$ has the same effect, since an increase in $b$ 
can be interpreted as if a bias was added to the input signal.
This is illustrated in Fig.~\ref{two_fig}b where the FHN neuron's 
spontaneous activity is displayed for different values of $a$ and $b$.

A marked ``ridge'' is present in the divergence surface in
Fig.~\ref{a-b_variations}a, indicating that there is a family of values of
the potential parameter $a$ and the bias parameter $b$ that would optimize
the ability of the modeled system to detect a (weak) sinusoidal signal.
The FHN neurons corresponding to these parameter values have the common
property that they fire only sparsely without the signal input but fire with
a significant intensity when the signal is present.
For parameter values outside the region under the ridge, the Kolmogorov
divergence, and associated performance, is uniformly lower.
The ``plateau'' on the left of the ridge is located above parameter values
for which the FHN neurons are very easily excited.
Given that the spike intensities of the FHN neurons corresponding to these
parameter values are roughly independent of the presence or absence of an
input signal, the presence of the plateau may seem counterintuitive.
However, the firing that takes place when an input signal is applied is much
more regular (since it is \textit{phase locked} to the signal) compared to
that taking place when the excitation is just noise.
Thus, the divergences corresponding to the systems for which the FHN parts
are easily excited are rather large but still clearly smaller than those
corresponding to the ridge.
In the former region of parameter values it is also possible that an applied
input signal decreases the firing rate since the noise-induced firing rate
can be larger than the rate given by a phase-locked spike train.
Consequently, even though the region of spontaneous firing yields rather
large divergences they are are clearly smaller than the divergences on the 
ridge.
The region of low divergences to the right of the ridge is generated by
parameter values corresponding to systems of FHN neurons that are very 
difficult to excite and hardly ever fire, even in the presence of an input 
signal.

Performing the same type of analysis on the system, but using the $\chi^2$ or
information divergence instead, yields qualitatively similar results, as seen
in Figs.~{\ref{a-b_variations}b,\ref{a-b_variations}c}.
Due to numerical problems it is hard to calculate the exact height of
the ridges, however, and we therefore limit the surfaces' heights in the
figures by truncating values above a certain threshold to the value of the
threshold.
Even though this prevents a precise estimation of the optimal combinations of
parameter values it allows the main objective to be fulfilled; to show the
existence of regions with (considerably) better performance in terms of
divergences than others.
For deflection ratios, on the other hand, the numerical problems are minor,
since they can be calculated without explicitly
calculating $p_0$ and $p_1$, which makes DRs more robust.
In Fig.~\ref{a-b_variations}d DRs for the output of the model are displayed.
Also for the DRs a ridge can be seen and the resulting set of optimal
values is similar to that for the divergences (though small changes in the
position of the ridge can be seen).
This qualitative behavior seen in all examples so far, with a (largely)
common region of optimal values, is recurrent in all our simulations
described in the following.
\subsection{Performance for a lower intensity level.}
In the previous section we described a simulation which was aimed at
investigating optimization of performance as a function of the potential 
parameter $a$ and the bias parameter $b$, in an otherwise fixed environment.
If we change the environment, new values of the parameters will emerge as
optimal.
For instance, if we lower the intensity level of the noise the location of
the ridge appearing in Fig.~\ref{a-b_variations}a will change, as seen in
Fig.~\ref{misc_variations}a.
Together, these two figures illustrate, moreover, that care must be exercised
when interpreting results of the stochastic resonance type
\cite{Gammaitoni_et_al} for neural processing systems:
For a \textit{fixed} pair of parameters values $a,b$, such as $a=0.6$ and 
$b=0.12$, the divergence can be higher for a larger noise level but the 
\textit{maximally achievable} divergence, obtained on the ridge in the two
figures, will be lower.
Hence, for a system where \textit{adaption} to environmental changes is
possible, a lower noise intensity is always better in our setting.
\subsection{Performance with respect to variation of $a$ and $\delta$.}
If we instead of varying the potential parameters $a,b$ vary the relaxation
parameter $\delta$ we get the result illustrated in
Fig.~\ref{misc_variations}b.
Also this divergence surface displays a marked ridge, similar to the one in 
Fig.~\ref{a-b_variations}a, indicating possible combinations of parameter
values for best performance.
\subsection{Performance for other input signal parameters.}
The ridges in the divergence surfaces discussed so far are only relevant for
the given input signal and if we change the input by e.g.\ altering the
amplitude or the frequency of the signal we get a different result.
Examples of this are shown in Fig.~\ref{misc_variations}c, where the
amplitude $A$ is set to 0.1, and in Fig.~\ref{misc_variations}d where the
angular frequency $\omega_0$ is set to 2.
Even though we still can see ridges in both cases they are different in shape
than the first one in Fig.~\ref{a-b_variations}a.
Obviously, the divergence decreases with decreased signal amplitude and the 
height of the ridge becomes lower in Fig.~\ref{misc_variations}c, but the
location of the ridge changes only slightly and it appears as if only a
slight change of optimal parameter values occurs.
When varying the frequency however, the ridge clearly moves to an entirely
new position and new parameter values render optimal performance.
\section{Discussion}
\label{Discussion}

\begin{figure}
  \center
  \begin{tabular}{cc}
  \includegraphics[width=5.5cm,height=5cm]{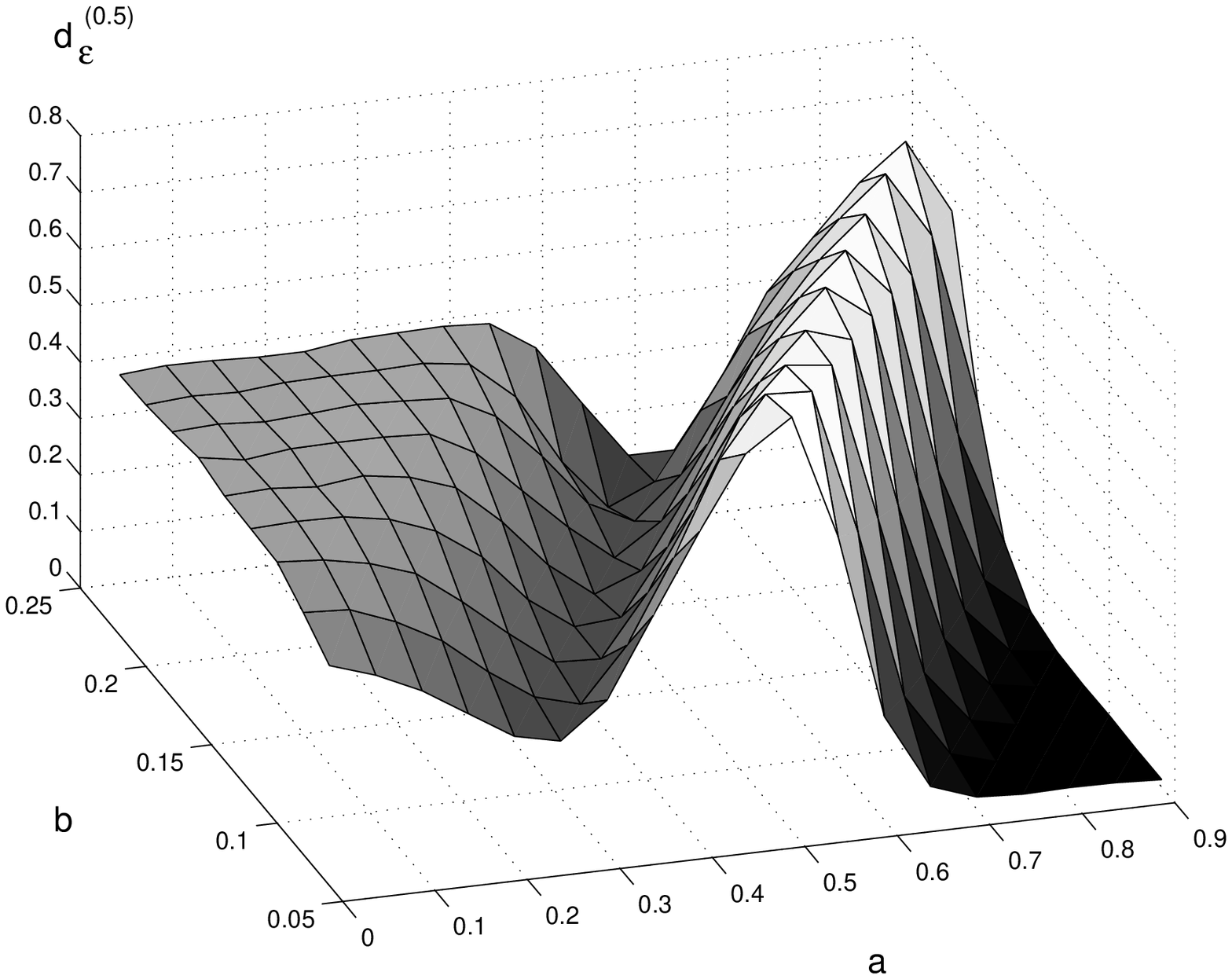} & 
  \includegraphics[width=5.5cm,height=5cm]{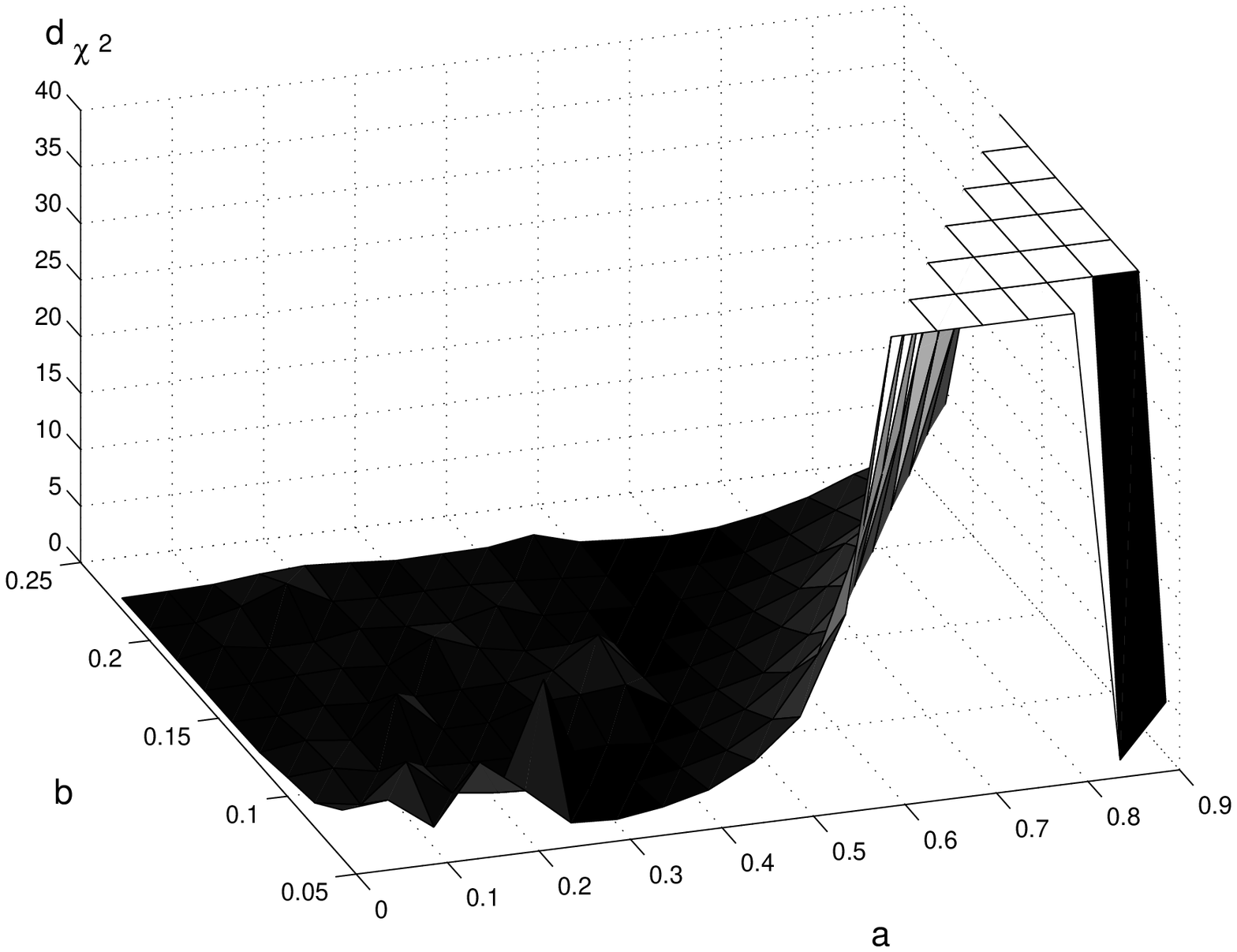} \\
  \includegraphics[width=5.5cm,height=5cm]{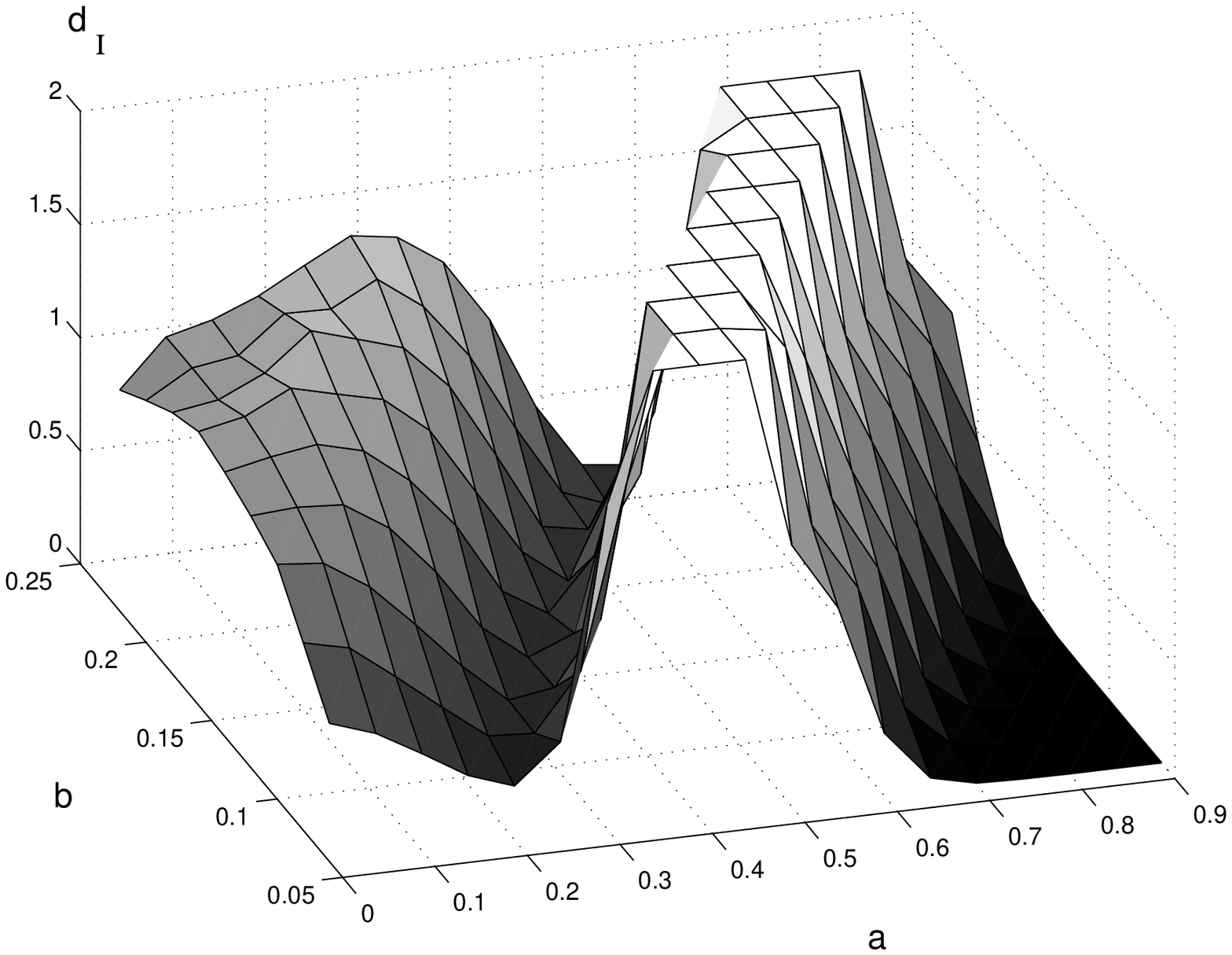} &
  \includegraphics[width=5.5cm,height=5cm]{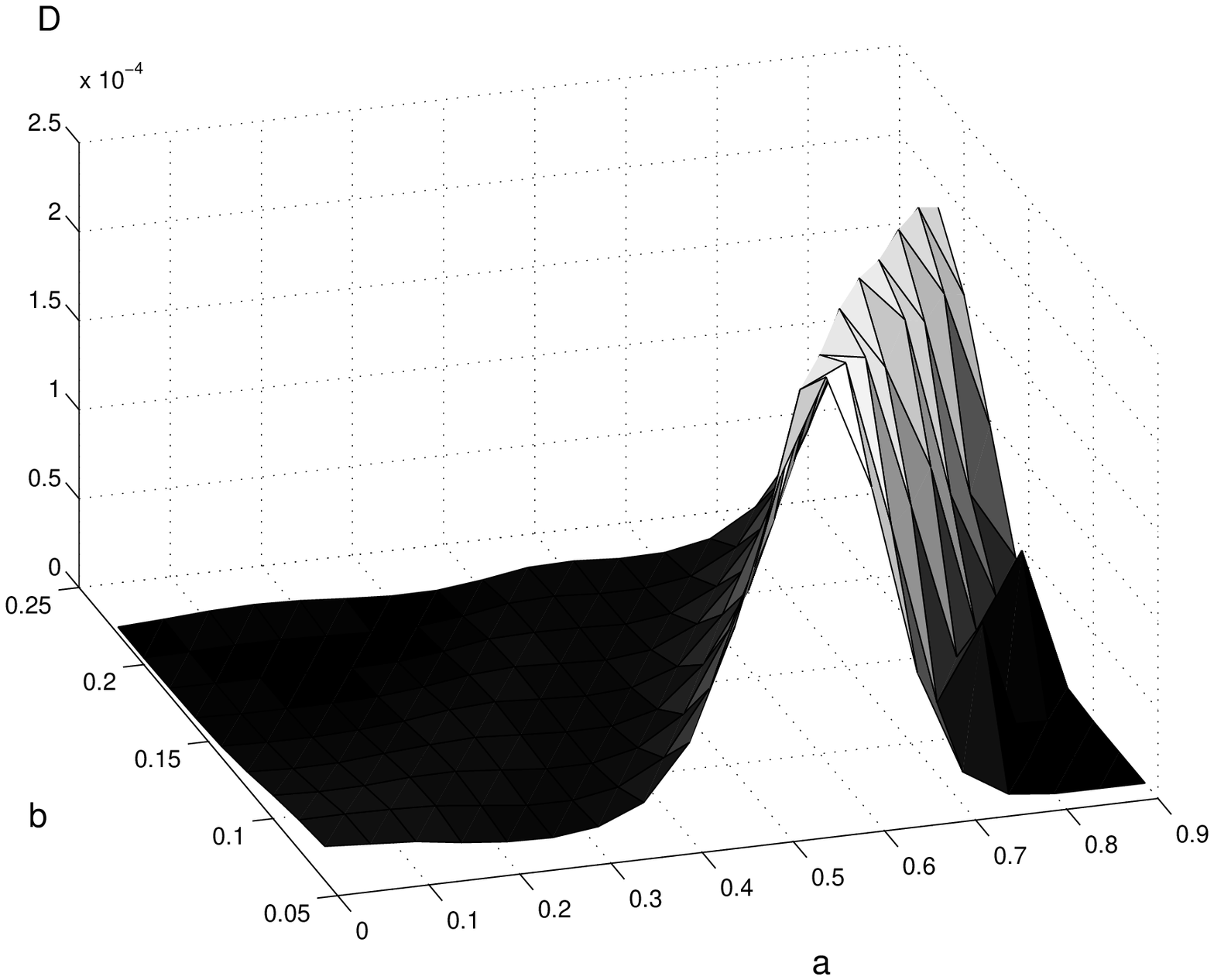}
  \end{tabular}
\caption{(a) Top left: The Kolmogorov divergence for $\delta=1$, 
             $\varepsilon = 0.005$, $A=0.2$, $\omega_0=8$, 
             $\sigma=100\sqrt{2\cdot 10^{-5}}$, $\lambda=100$, $\alpha=10$, 
             $\beta=100$, $x_0=0.25$, $L=1.5$, $R=0.2$, $\tau=50$, 
             $\gamma=0.5$, $q=0.5$ and different values of the potential 
             parameter $a$ and the bias parameter $b$.
         (b) Top right: The $\chi^2$ divergence for different values of the
             potential parameter $a$ and the bias parameter $b$ when the 
             other parameter values are the same as in 
             Fig~\ref{a-b_variations}a.
             Due to the unreliability for high values of the divergence no
             value above 40 has been plotted. (Eventually the
             $\chi^2$-divergence decreases to zero, when $a$ becomes
             sufficiently large, since then there will be almost no spikes
             generated.)
         (c) Bottom left: The information divergence for different
             values of the potential parameter $a$ and the bias parameter $b$ 
             when the other parameter values are the same as in 
             Fig~\ref{a-b_variations}a.
             Due to the unreliability for high values of the divergence no
             value above 3 has been plotted.
         (d) Bottom right: The deflection ratio for different
             values of the potential parameter $a$ and the bias parameter
             $b$ when the other parameter values are the same as in 
             Fig~\ref{a-b_variations}a.}
\label{a-b_variations}
\end{figure}

\begin{figure}
  \center
  \begin{tabular}{cc}
  \includegraphics[width=5.5cm,height=5cm]{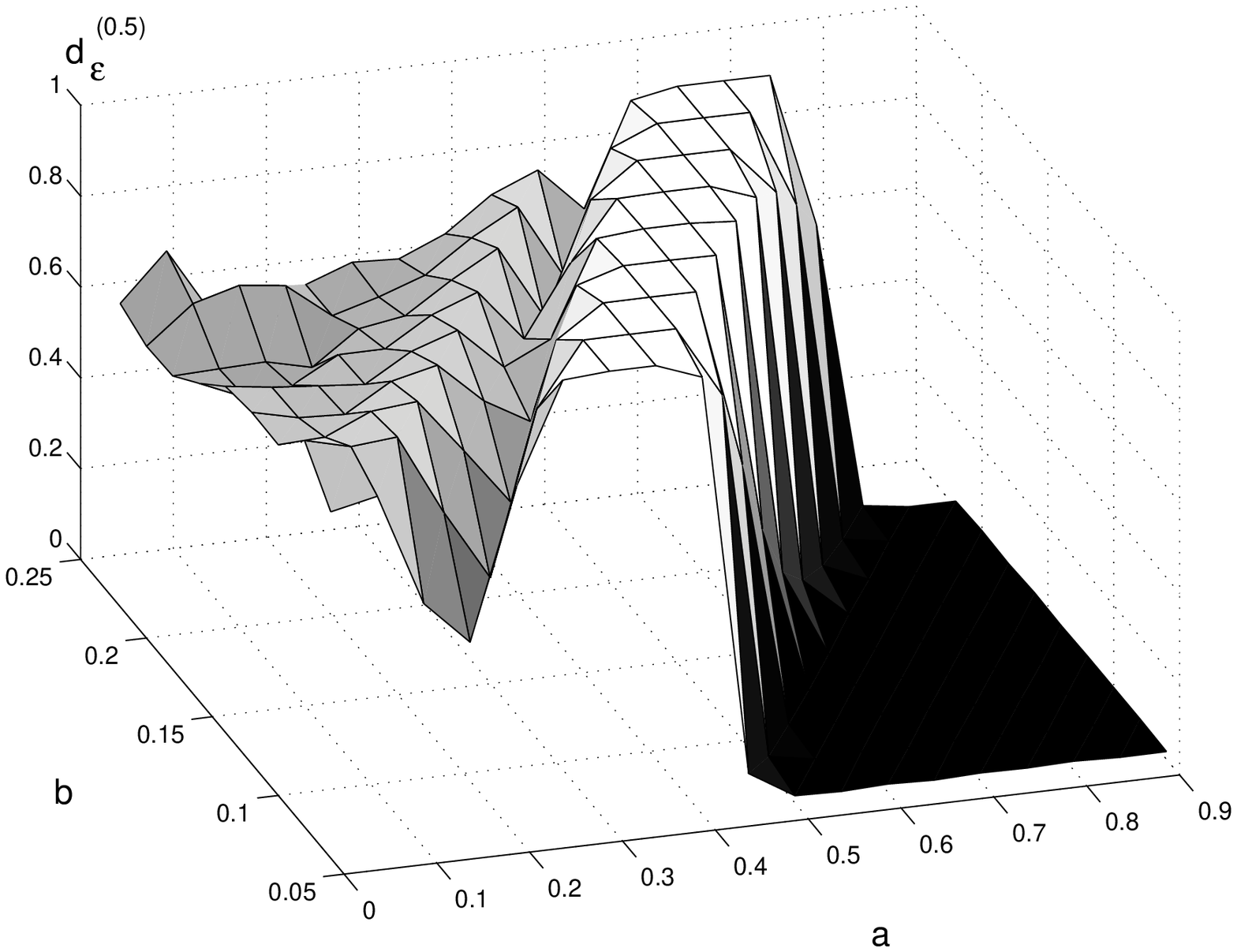} &
  %\label{ab_D}
  \includegraphics[width=5.5cm,height=5cm]{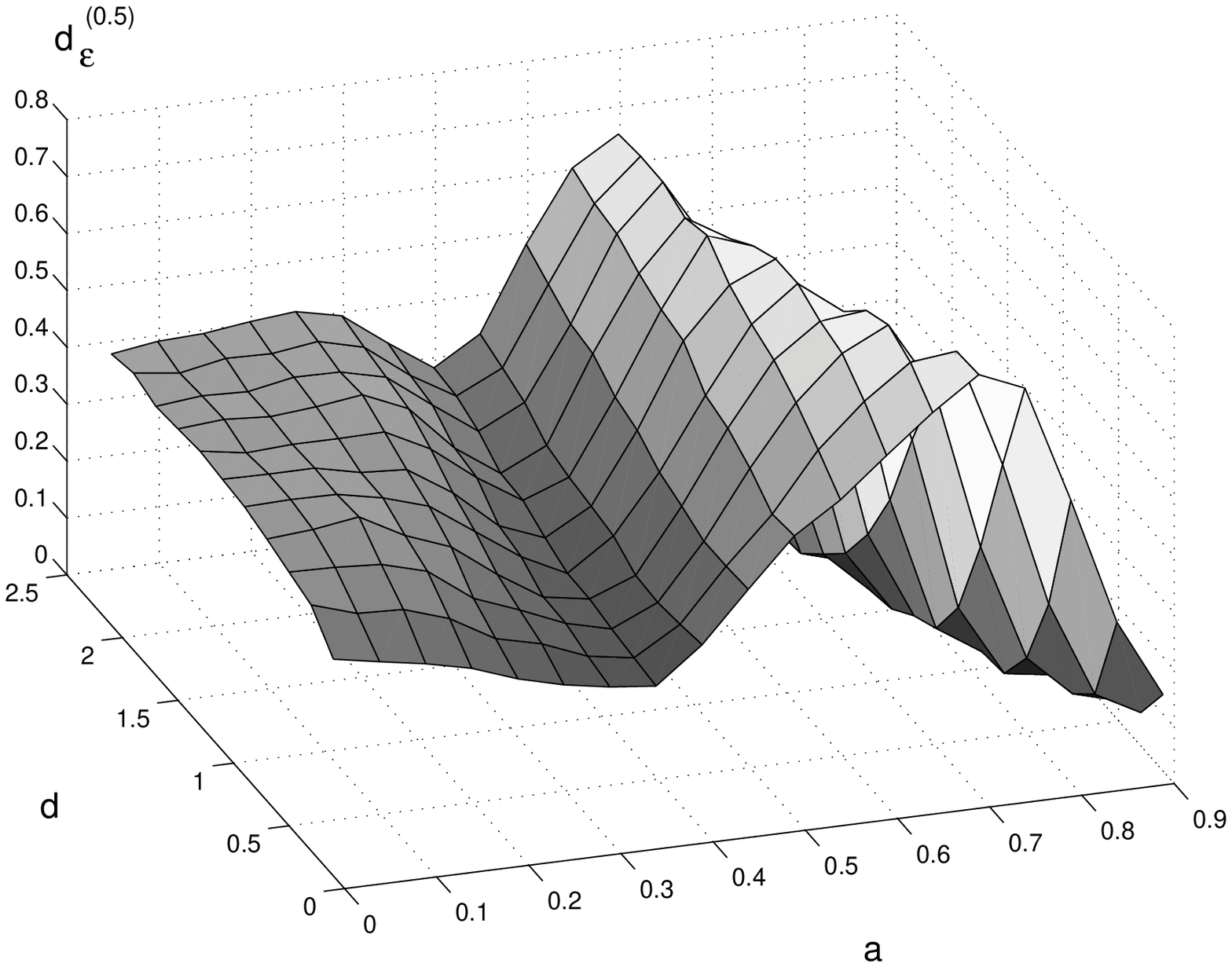} \\
  %\label{ad_var}
  \includegraphics[width=5.5cm,height=5cm]{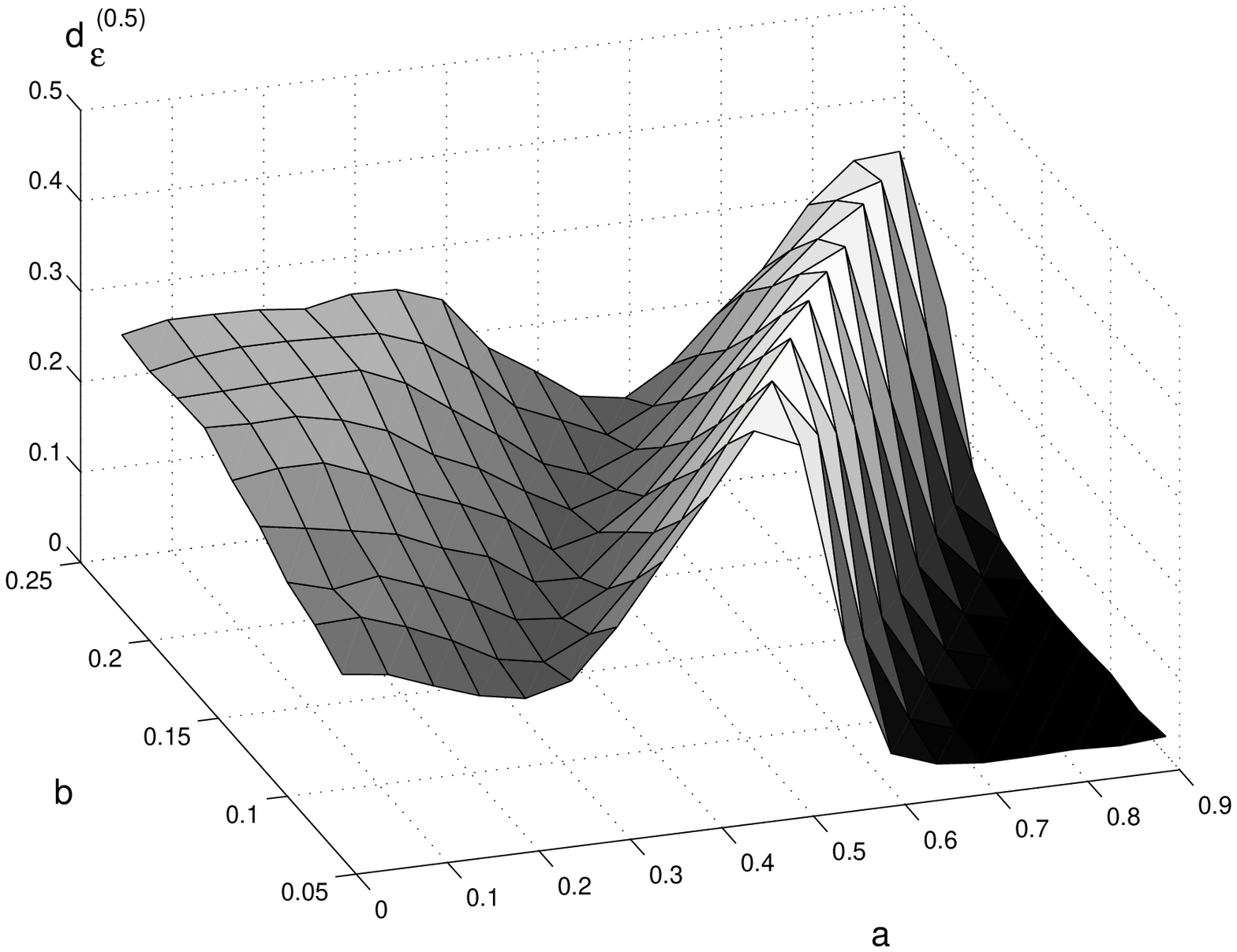} &
  %\label{ab_r}
  \includegraphics[width=5.5cm,height=5cm]{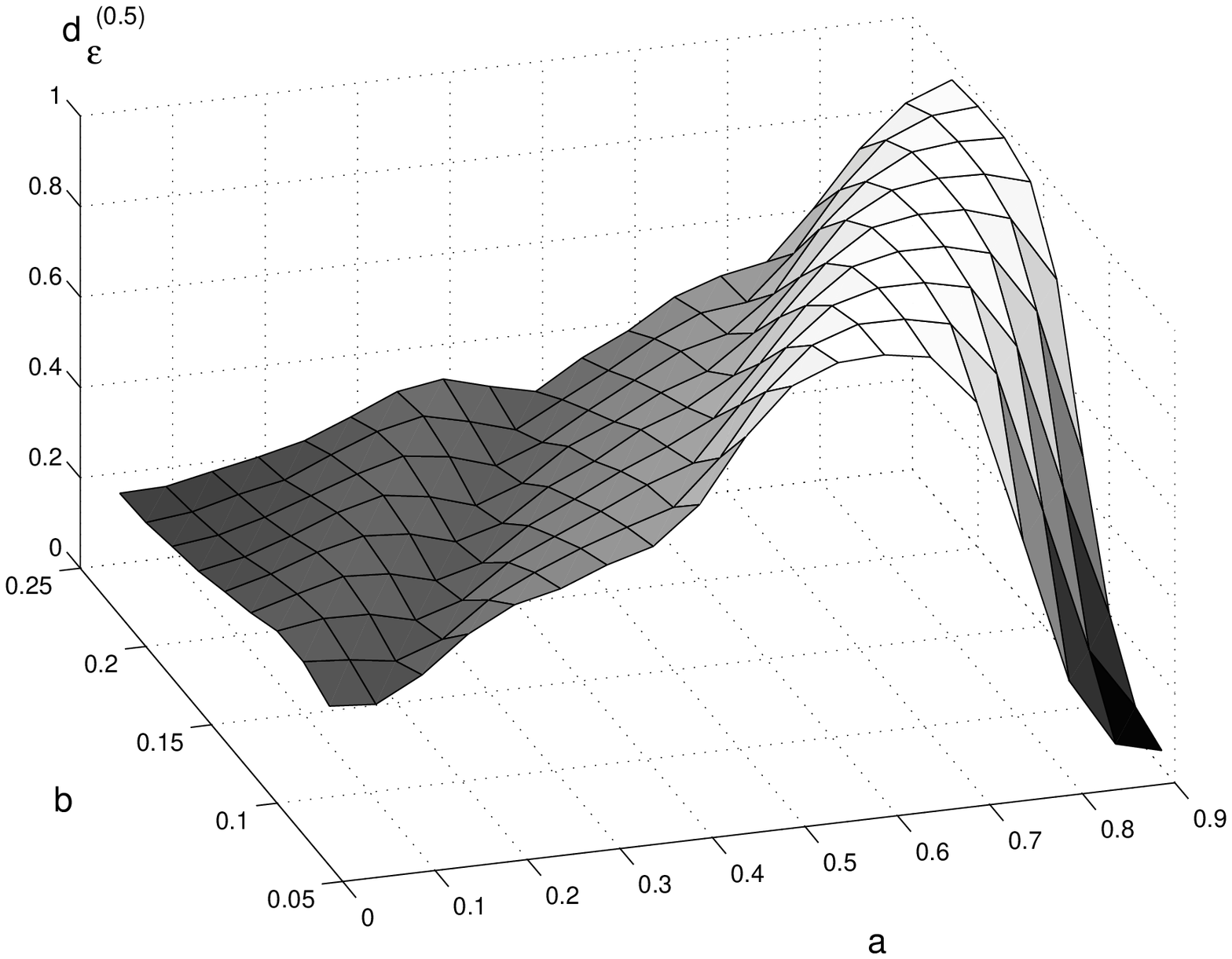}
  %\label{ab_beta}
  \end{tabular}
\caption{(a) Top left: The Kolmogorov divergence for different values of the
             potential parameter $a$ and the bias parameter $b$, when the 
             other parameter values are the same as in 
             Fig~\ref{a-b_variations}a except for the noise intensity, which
             is lower ($\sigma=100\sqrt{2\cdot10^{-6}}$).
         (b) Top right: The Kolmogorov divergence for different values of the
             potential parameter $a$ and the relaxation parameter $\delta$
             for $b=0.12$ and with the other parameter values as 
             in Fig~\ref{a-b_variations}a.
         (c) Bottom left: The Kolmogorov divergence for different values of
             the potential parameter $a$ and the bias parameter $b$ when the 
             other parameter values are the same as in 
             Fig~\ref{a-b_variations}a except for the signal amplitude, 
             which is lower ($A=0.1$).
         (d) Bottom right: The Kolmogorov divergence for different values of
             the potential parameter $a$ and the bias parameter $b$ when the 
             other parameter values are the same as in 
             Fig~\ref{a-b_variations}a except for the frequency, which is 
             lower ($\omega_0=2$).}
\label{misc_variations}
\end{figure}

We have described a method for analyzing the information processing
capability in the primary part of the mammalian auditory nervous system using
fundamental statistical and information theoretical performance criteria,
quantitatively expressed by $\phi$-divergences.
Our premise has been that, since these criteria are highly relevant for the
processing taking place in this system, the non-existence of well defined
global maxima of these criteria occuring in the interior of regions of
feasible system parameters would suggest incompleteness or incorrectness of
the overall model.
(Loosely speaking, one can argue that such global \textit{interior} maxima
must exist for the ``right'' criteria in a ``correct'' model since otherwise
parameters would have to be set at boundaries in order to achieve optimal
behavior.
Parameters at boundaries would favor \textit{structural} change by evolution
until only interior optima occur, whereby the ``drive'' for structural
change ceases).
One instance of this point is that without taking into account the
synaptic plasticity, it can be shown that the divergence surfaces will have a
qualitatively different shape, with an additional ridge that, at least
partly, will yield optimal parameter values that are unphysical.
However, the observed ``ridges'' in the divergence and deflection surfaces
in Figs.~\ref{a-b_variations},\ref{misc_variations} indeed allow for
optimization of performance by taking parameter values in the interior of the
domain of values that have physical significance. 
Since the model is based on fairly standard and well accepted components
(e.g.\ the FHN model), which we feel capture the essential mechanisms
involved in the information processing considered here, we believe that the
results in fact can be interpreted as a quantitative indication of how some
of the parameters in the auditory system presumably must be set.
In particular this applies to the \textit{quiescent firing rate} (QFR) which,
in real systems under this assumption, must take values (as a result of
evolution) near those that correspond to the maxima of the performance
measures considered here.
Verifying this is a topic for further research, however.

The conclusion about the QFR is based on the qualitative observation that all
the ``ridges'' appearing in the divergence and deflection surfaces correspond
to parameter values that lie a certain ``thin'' or ``manifold
like'' set in parameter space.
A closer examination of this set shows that the combinations of parameter
values that correspond to e.g.\ the ridge in Fig.~\ref{a-b_variations}a
describe systems that have virtually the same firing intensity in the absence
of an external signal, i.e.\ virtually the same QFR.
Since this specific QFR also is common for all optimal values of parameter 
combinations corresponding to the ridges in
Figs.~{\ref{misc_variations}a,\ref{misc_variations}b}, and in all other
simulations that we have tried with the same input signal, this strongly
suggests a connection between the QFR and the information processing
performance of the system.
Further evidence supporting this hypothesis can be seen in 
Figs.~{\ref{misc_variations}c,\ref{misc_variations}d} which show that the
optimal QFR, and thereby the optimal set of parameters, is very little
affected by a change in our (weak) signal amplitude but changes considerably 
with the applied frequency.
This reflects well the frequency division of sound performed in the inner
ear, as discussed in Section \ref{Inf_proc}.
A more detailed investigation of the frequency dependence also shows that 
the optimal QFR in our model increases with increasing frequency.
Even though existing real data is inconclusive on this point, Kiang's
classical data \cite{Kiang} can be interpreted to support the hypothesis that
such a frequency dependence exists.
However, experiments are needed to resolve the issue.
Finally we point out that even though the location of the ridge in
e.g.\
Figs.~{\ref{a-b_variations}a,\ref{a-b_variations}b,\ref{a-b_variations}c} is
largely the same it does vary slightly depending on which divergence or
deflection is considered, which is to be expected since these performance
measures are not identical.
In particular, the $\chi^2$-divergence in Fig.~\ref{a-b_variations}b can, as
explained in Sec.\ \ref{diverg_sect}, be considered to be a first order
approximation of the information divergence in Fig.~\ref{a-b_variations}c.

All constants in our model have been chosen in order to produce as 
realistic data as possible.
The choices are not critical though, since in most of the simulations
where the values of the constants are varied (in a reasonable large 
interval) the results are qualitatively invariant.
Our approach therefore offers a new qualitative, and possibly also a
quantitative, explanation of the different levels of QFRs observed in 
the auditory nerve.

\section{Acknowledgements}
The authors would like to acknowledge fruitful discussions with
Prof.\ A.~Longtin of Ottawa Univ.\ which led to improvement of the results in
several aspects and to Dr.\ A.\ Bulsara of SPAWAR SSC, San Diego, CA, for
many insightful suggestions which clarified the presentation of the material.
MFK would also like to thank Prof.\ Longtin for his hospitality
during a visit in Ottawa.
%
%\section{References}
%

%
%\section{Figure Legends}
%\section{Tables}
\section{Figures}

%
%\section{Appendices}
%

\end{article}
\end{document}